\documentclass[aps,pre,reprint,longbibliography,floatfix,superscriptaddress]{revtex4-2}
\usepackage{graphicx}
\usepackage{amsmath,amssymb,amsfonts}
\usepackage{hyperref}

\begin{document}

\title[Connectedness percolation in the RSA packings of elongated particles]
{Connectedness percolation in the random sequential adsorption packings of elongated particles}
\author{Nikolai I. Lebovka}
\email[Corresponding author: ]{lebovka@gmail.com}
\affiliation{Laboratory of Physical Chemistry of Disperse Minerals, F. D. Ovcharenko Institute of Biocolloidal Chemistry, NAS of Ukraine, Kyiv 03142, Ukraine}
\affiliation{Department of Physics, Taras Shevchenko Kiev National University, Kyiv 01033, Ukraine}

\author{Mykhailo O. Tatochenko}
\email{tatochenkomihail@gmail.com}

\author{Nikolai V. Vygornitskii}
\email{vygornv@gmail.com}
\affiliation{Laboratory of Physical Chemistry of Disperse Minerals, F. D. Ovcharenko Institute of Biocolloidal Chemistry, NAS of Ukraine, Kyiv 03142, Ukraine}

\author{Andrei V. Eserkepov}
\email{dantealigjery49@gmail.com}

\author{Renat K. Akhunzhanov}
\email{akhunzha@mail.ru}

\author{Yuri Yu. Tarasevich}
\email[Corresponding author: ]{tarasevich@asu.edu.ru}
\affiliation{Laboratory of Mathematical Modeling, Astrakhan State University, Astrakhan 414056, Russia}

\date{\today}

\begin{abstract}
Connectedness percolation phenomena in the two-dimensional packing of elongated particles (discorectangles) were studied numerically. The packings were produced using random sequential adsorption (RSA) off-lattice models with preferential orientations of the particles along a given direction. The partial ordering was characterized by the order parameter $S$, with $S=0$ for completely disordered films (random orientation of particles) and $S=1$ for completely aligned particles along the horizontal direction $x$. The aspect ratio (length-to-width ratio) of the particles was varied within the range $\varepsilon \in [1;100]$. Analysis of connectivity was performed assuming a core--shell structure of the particles. The value of $S$ affected the structure of the packings, the formation of long-range connectivity and the behavior of the electrical conductivity. The effects can be explained by taking accounting of the competition between the particles' orientational degrees of freedom and excluded volume effects.
For aligned deposition, anisotropy in the electrical conductivity was observed with the values along the alignment direction, $\sigma_x$, being larger than the values in the perpendicular direction, $\sigma_y$. Anisotropy in the localization of the percolation threshold was also observed in finite sized packings, but it disappeared in the limit of infinitely large systems.
\end{abstract}

\maketitle

\section{Introduction\label{sec:intro}}

The random packing of elongated particles onto a plane is a challenging problem that has been the ongoing focus of many researchers. The particle shape may affect the packing characteristics (e.g., packing density and coordination numbers)~\cite{Zou1996,Guises2009,Kyrylyuk2011}, the aggregation~\cite{Kwan2001}, and the gravity- and vibration-induced segregation~\cite{Abreu2003}. A lot of interest in such systems continues to be stimulated by practical problems related to the  preparation of advanced materials~\cite{Bokobza2019,Yang2019} and composite films~\cite{Hirotani2019,Tiginyanu2019}, filled with elongated nanoparticles, e.g., carbon nanotubes~\cite{Pampaloni2019} and silicate platelets~\cite{Lebovka2019Springer}.

For the simulation of random packings, random sequential adsorption (RSA) models~\cite{Evans1993,Adamczyk2012} are frequently used. In such models, the particles are deposited randomly and sequentially onto a two-dimensional (2D) substrate without overlapping. At  the so-called ``jamming limit'', where $\varphi_\text{j}$ is the saturated coverage concentration,   no more particles can be adsorbed and the deposition process terminates. The problems related to the kinetics of 2D RSA, the jamming limit, and the asymptotic behavior of RSA deposition for elongated particles (ellipses, rectangles, discorectangles, and needles) were discussed in detail~\cite{Talbot1989,Tarjus1991,Viot1992,Viot1992a,Ricci1992,Talbot2000}.  The  saturated 2D RSA packings for different particle shapes, including disks~\cite{Feder1980}, ellipses~\cite{Talbot1989,Sherwood1990}, squares~\cite{Viot1990}, rectangles~\cite{Vigil1989,Vigil1990}, discorectangles~\cite{Haiduk2018,Lebovka2020b},  polygons~\cite{Ciesla2014}, sphere dimers, sphere polymers, $k$-mers and extended objects~\cite{Perino2017,Budinski-Petkovic2016,Lebovka2020a}, and other shapes~\cite{Ciesla2013,Ciesla2013a,Ciesla2015} have been studied in detail. Particularly, for very elongated unoriented particles the saturation coverage gone to zero when the aspect ratio becomes infinite~\cite{Viot1992,Viot1992a}. Moreover, the non-monotonic dependencies of the values of $\varphi_\text{j}$ versus the aspect ratio,  $\varepsilon$, have been observed. Similar dependencies have also been observed for saturated RSA packings of elongated particles in one-dimensional (1D)~\cite{Baule2017,Ciesla2020,Lebovka2020} and three-dimensional (3D)~\cite{Donev2004,Chaikin2006,Gan2020a} systems. The appearance of maximums of the jamming concentration can be explained by a competition between the effects of orientational degrees of freedom and excluded volume effects~\cite{Donev2004}.

The formation of long-range connectivity is the primary issue to be solved for better understanding of the percolation phenomena of core--shell anisotropic particles in random packings. Core--shell composite particles consist of an inner layer of one material (the core) and an outer layer of another material (the shell). Core--shell particles have already demonstrated promising applications in electrochemical, optical, wearable and gas adsorptive sensors~\cite{Kalambate2019}, electrode materials~\cite{Ho2019},  polymeric composites~\cite{Ryu2018} and drug delivery applications~\cite{Schmitt2020}. The practical significance of the problem is also related to a need to obtain a description of the behavior of the electrical conductivity of composites filled with elongated core--shell particles, e.g., carbon nanotubes and fibers, metallic nanorods and nanocables, and other core--shell particulates~\cite{Balberg1987,Berhan2007,White2010,Nan2010,Panda2014,Tomylko2015,Ryu2018,Chen2018,Xu2020,Islam2020,Balberg2020}. In general, the inner material can be covered partially or fully by a single or multiple outer layers. By regulation of the shell properties, materials with enhanced optical, electrical, or magnetic characteristics, and improved thermal stability or dispersibility can be obtained. For particles with core--shell structures, their resulting electrical conductivity can reflect the effects of particle ordering, packing, connectivity rules and the intrinsic properties of the cores, the matrix, and the interface between the particles and the matrix (shells).

In this paper, we shall concentrate on the percolation effects in 2D RSA packings of discorectangles. A hard core---soft shell structure of particles was assumed and anisotropic packing with preferential orientation of the particles along a given direction were considered. The effects of the particle aspect ratios, orientation ordering, and packing fraction on the electrical conductivity of the packings together with the critical thickness of the shells required for a spanning path through the system were evaluated. The rest of the paper is organized as follows. In Sec.~\ref{sec:methods}, the technical details of the simulations are described and all necessary quantities are defined. In order to provide a better understanding in respect of the precision of the calculations a range of some test results are also given. Section~\ref{sec:results} presents our principal findings and discussions. Finally, Section~\ref{sec:conclusion} summarizes our findings.

\section{Computational model\label{sec:methods}}
A discorectangle is a rectangle with semicircles at a pair of opposite sides. The discorectangles were randomly and sequentially deposited until they reached the saturated coverage concentration $\varphi_\text{j}$. An optimized RSA algorithm, based on the tracking of local regions, was used~\cite{Haiduk2018,Lebovka2020b}. The aspect ratio (length-to-width ratio) was defined as $\varepsilon=l/d$, where $l$ is the length of the particle and $d$ is its width. Discorectangles with $\varepsilon \in [1;100]$ were considered.

The degree of orientation was characterized by the order parameter defined as
\begin{equation}\label{eq:S}
  S = \left\langle \cos 2\theta  \right\rangle,
\end{equation}
where $\langle\cdot\rangle$ denotes the average, $\theta$ is the angle between the long axis of the particle and the direction of the preferred orientation of the particles ($x$ direction).

For generation of the aligned packings, the orientations of the deposited particles were  selected to be uniformly distributed within some interval such that $-\theta_\text{m} \leqslant \theta \leqslant \theta_\text{m}$, where $\theta_\text{m} \leqslant \pi/2$~\cite{Balberg1983}. For the selected model of deposition~\cite{Balberg1983} the order parameter was calculated as~\cite{Lebovka2019}
\begin{equation}\label{eq:S0}
S = \frac{\sin 2\theta_\text{m}}{2\theta_\text{m}}.
\end{equation}

\begin{figure*}[!htbp]
\centering
\includegraphics[width=0.6\textwidth]{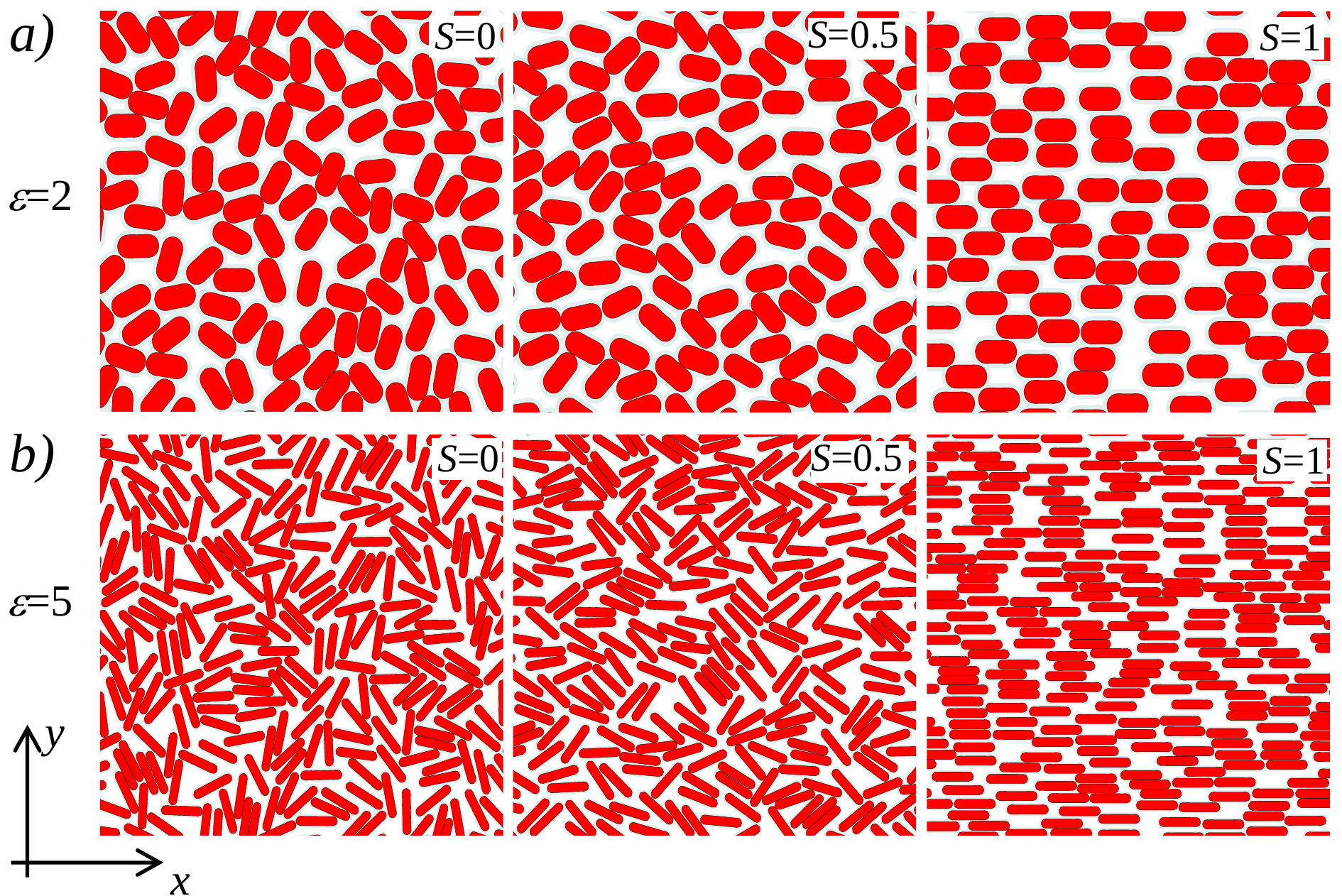}
\caption{Examples of RSA packings in the jamming state for discorectangles with aspect ratios $\varepsilon = 2$ (a); and $\varepsilon = 5$ (b); and at different values of the order parameters:  $S=0$ (random orientation), $S=0.5$ (partial orientation) and $S=1$ (complete alignment along the horizontal direction $x$).\label{fig:f01}}
\end{figure*}

Figure~\ref{fig:f01} shows examples of the packing patterns in the jamming state for discorectangles with aspect ratios $\varepsilon = 2$, (a); and  $\varepsilon = 5$, (b). For random orientation of particles  ($\theta_\text{m} = \pi/2$) we have $S=0$ and for complete alignment of particles along the horizontal direction $x$ ($\theta_\text{m} = 0$) we have $S=1$. For intermediate values $0<S<1$ during the deposition, some particle orientations may be rejected and the real order parameter in the deposit may differ from the preassigned value~\cite{Lebovka2011,Lebovka2020b}.

\begin{figure}[!htbp]
	\centering	
\includegraphics[width=\columnwidth]{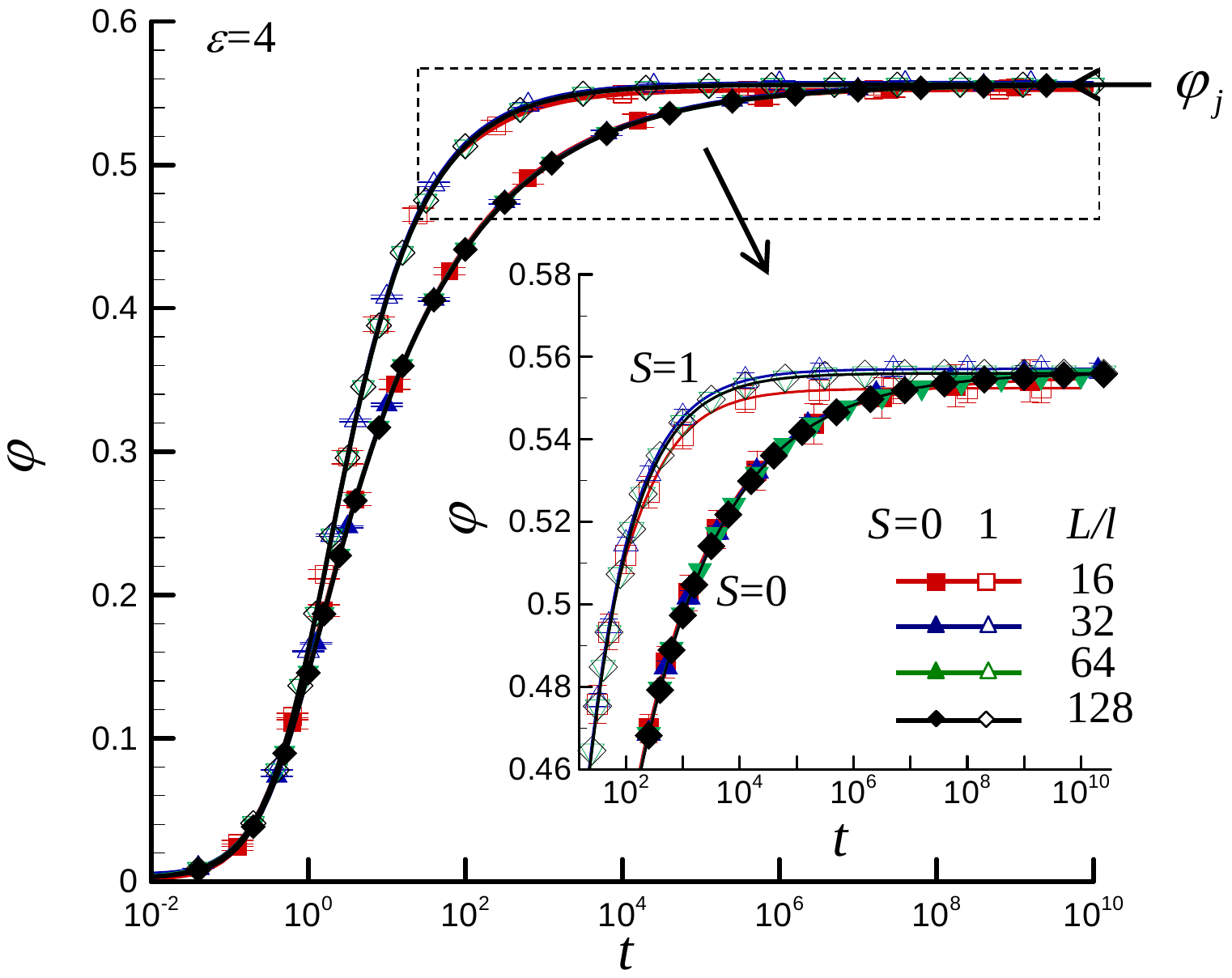}\\
	\caption{Coverage concentration $\varphi$ versus the deposition time, $t$, for the RSA packing of random ($S=0$)  and perfectly aligned ($S=1$) discorectangles with aspect ratio $\varepsilon=4$ at different values of $L/l$. Here, $\varphi_\text{j}$ is the jamming coverage. Inset shows an  enlarged portion of the $\varphi (t)$ plot near the saturation concentration.
 \label{fig:f02} }
\end{figure}

The dimensions of the system under consideration were $L$ along both the horizontal ($x$) and the vertical ($y$) axes, and periodic boundary conditions were applied in both directions. The time was measured using dimensionless time units, $t=n/L^2$, where $n$ is the number of deposition attempts. Figure~\ref{fig:f02} shows examples of the coverage concentration $\varphi$ versus the deposition time, $t$ , for the RSA packing of random ($S=0$)  and perfectly aligned ($S=1$) discorectangles with aspect ratio $\varepsilon=4$ at different values of $L/l$. Similar dependencies were observed for other values of $S$ and $\varepsilon$. The scaling tests with $L/l = 16, 32, 64$, and $128$ evidenced the good convergence of the data at $L/l \geqslant 32$. In the present work, the majority of calculations were performed using $L = 32l$ and the jamming coverage was assumed to be reached after a deposition time of $t=L^2 \times 10^{10}$.

\begin{figure*}[!htbp]
	\centering
	\includegraphics[width=\textwidth]{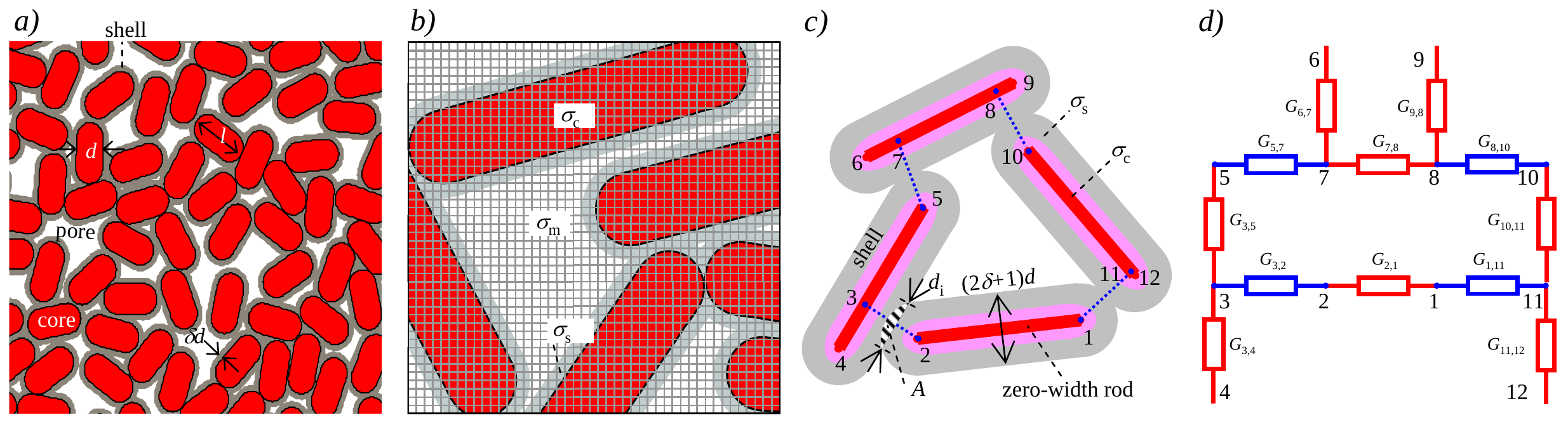}
	\caption{Approaches to description of the connectivity analysis (a) and calculation of electrical conductivity (b) of the RSA packing of discorectangles on a 2D substrate. A core--shell structure of the particles was assumed. Intersections of the particle cores were forbidden. For the connectivity analysis, each particle was assumed to be covered by a soft (penetrable) shell of thickness $\delta d$. To calculate the electrical conductivity, $\sigma$, a discretization approach with a supporting mesh was used. The mesh cells with centers located at the cores, shells, or pores parts were assumed to have electrical conductivity of $\sigma_\text{c}$, $\sigma_\text{s}$ and $\sigma_\text{m}$, respectively.  (c) For larger values of the aspect ratio (slender-rod limit),  discorectangles were treated as zero-width rods with the electrical conductivity $\sigma_\text{c}$.
(d) Example of a transformation of slender rods into a RRN. \label{fig:f03}}
\end{figure*}

The analysis of the connectivity was performed assuming a core--shell structure of the particles, with particle having an outer shell of thickness $\delta d$ (Fig.~\ref{fig:f03}a). Any two particles were assumed to be connected when the minimal distance between their hard cores did not exceed the value of $\delta d$. The connectivity analysis was carried out using a list of near-neighbor particles~\cite{Marck1997}. The minimum (critical) value of the relative outer shell thickness, $\delta_\text{c}$, (hereinafter, the shell thickness) required for the formation of spanning clusters in the $x$ or $y$ direction, was evaluated using the Hoshen---Kopelman algorithm~\cite{Hoshen1976}.

To calculate the electrical conductivity, $\sigma$, two approaches was used. Within the first one (m-model), the 2D plane was covered by a supporting square mesh of size $m \times m$ (Fig.~\ref{fig:f03}b). The mesh cells with centers located at the core, shell, or pore parts were assumed to have electrical conductivities of $\sigma_\text{c}$, $\sigma_\text{s}$ and $\sigma_\text{m}$, respectively. Then each cell was associated with a set of four resistors and the system was transformed into a random resistor network (RRN) ) (for more details see Appendix~\ref{sec:appendixmmodel}). Note that calculations at large values of $m$ provided better accuracy, but required significantly more computing resources. Therefore, the effects of the values $m$ ($m = 1024, 2048, 4096$) on the calculated values of $\sigma$ were also checked in some calculations. This approach has been used for the values of the aspect ratios up to~20. To calculate the electrical conductivity of the RRN the Frank---Lobb algorithm
based on the Y-$\triangle$ transformation was applied~\cite{Frank1988}. More detailed information on the calculation of the electrical conductivity can be found elsewhere~\cite{Tarasevich2018,Tarasevich2018a}.

For larger values of the aspect ratio (slender-rod limit), other approach (t-model) was used. The electrical conductivity of the substrate was ignored ($\sigma_\text{m}=0$). Within this approach, discorectangles were treated as zero-width rods with the electrical conductivity  $\sigma_\text{c}$. The electrical conductance between any two points (say, $i$ and $j$) belonging to the same rod is inverse proportional to the distance $l_{i,j}$between these point (see, e.g., \cite{Tarasevich2019JAPa,Tarasevich2019JAPb}). The electrical conductivity between any two rods with overlapping shells is proportional of  the width of the conduction channel (maximal width of the overlapping) and inverse proportional to its length (the effective distance between their cores) $G_{ij}^\text{s} = \sigma_\text{s} d_\text{i}/ l_\text{e}$. The effective distance may be estimated as
$$
l_\text{e} = \frac{2\delta d d_\text{i} - A}{d_\text{i}},
$$
where $A$ is the area of the overlapping shells and $d_\text{i}$ is the distance between the two intersection points of the outer perimeters of these shells (see Fig.~\ref{fig:f03}c,  Fig.~\ref{fig:f03}d, and Appendix~\ref{sec:appendix} for details of overlapping calculations). In the particular case of parallel or perpendicular rectangles with the core--shell structure, this approach provides exact values of the electrical conductance. In the case of arbitrary oriented discorectangles with the core--shell structure, this approach provides approximate values of the electrical conductance.

Then, Kirchhoff's current law was applied to each junction, and Ohm's law used for each circuit between two junctions. The resulting set of equations was solved to find the total conductance of the RRN. More detailed information on the calculation of the electrical conductivity can be found elsewhere~\cite{Tarasevich2019JAPa,Tarasevich2019JAPb}.

Large contrasts in electrical conductivities were assumed, $\sigma_\text{c} \gg\sigma_\text{s} \gg \sigma_\text{m}$. We let $\sigma_\text{c}=10^{12}, \sigma_\text{s}=10^6$ and $\sigma_\text{m}=1$ in arbitrary units. In this case, resistance of shells give the main contribution in the electrical resistance of the system under consideration, while resistance of cores has negligible contribution. For each given value of $\varepsilon$ and $S$, the computer experiments were repeated using from 10 to 1000 independent runs. The error bars in the figures correspond to the standard deviations of the means. When not shown explicitly, they are of the order of the marker size.

\section{Results and Discussion\label{sec:results}}

\subsection{Connectivity}
For a discorectangle, the critical shell thicknesses $\delta_{c,x}$ and $\delta_{c,y}$ correspond to the formation of percolation clusters in the $x$ and $y$ direction, respectively. For isotropic system with $S=0$, the values of $\delta_{c,x}$ and $\delta_{c,y}$ coincide, i.e., $\delta_{c,x}=\delta_{c,y}$. For anisotropic systems with $S\neq 0$, these values may be different. At a fixed value of shell thicknesses, $\delta$, the critical coverages $\varphi_{c,x}$ and $\varphi_{c,y}$, required for the formation of percolation clusters in the $x$ and $y$ directions, respectively, can be also defined.
\begin{figure}[!htbp]
	\centering
	\includegraphics[width=\columnwidth]{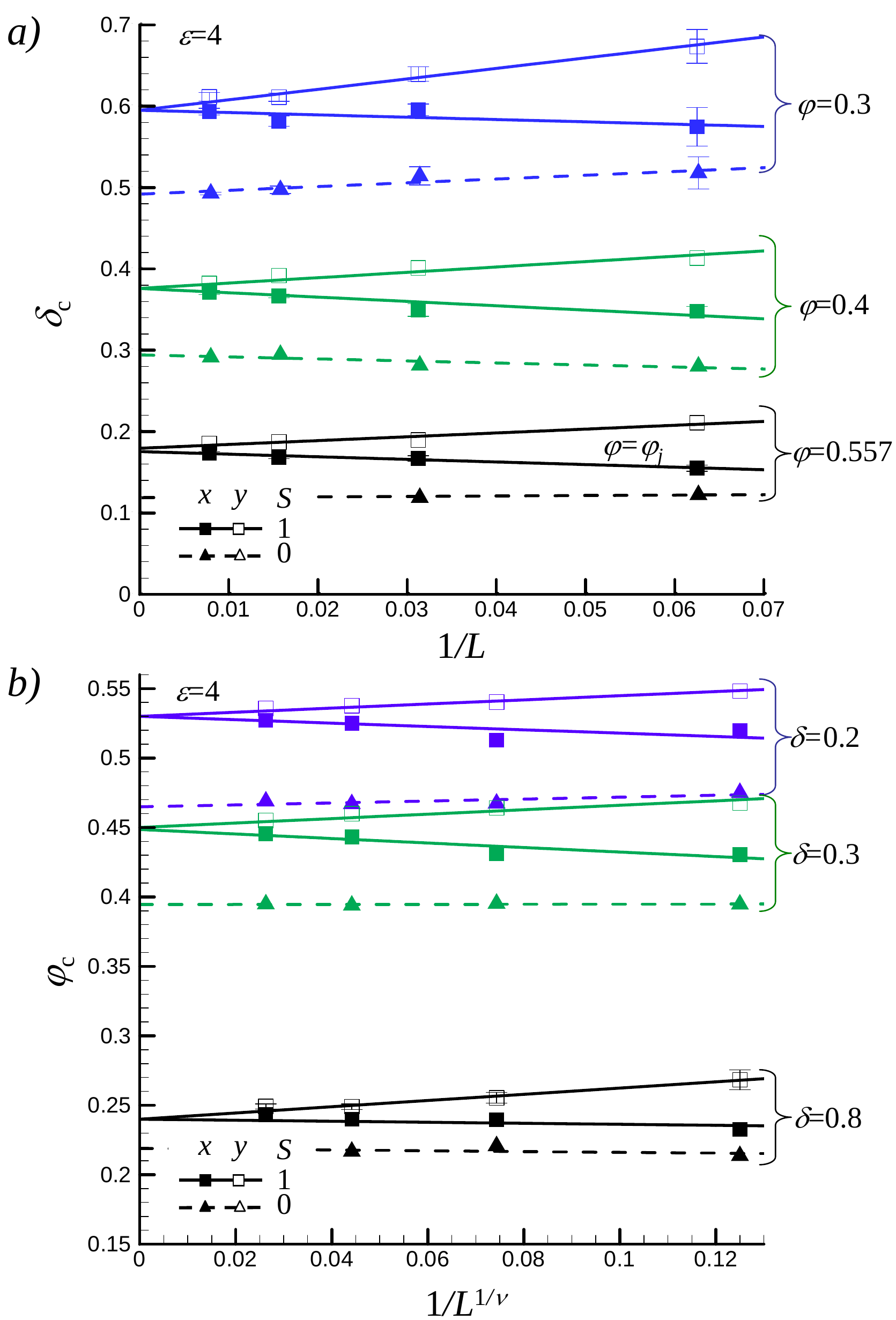}
	\caption{Scaling dependencies of the critical shell thickness $\delta_\text{c}$ at different values of particle coverage, $\varphi_{c}$, (a) and of the critical particle coverage $\varphi_\text{c}$ at different fixed values of shell thickness, $\delta$ (b). The data are presented for an aspect ratio of $\varepsilon=4$ for completely disordered ($S=0$, dashed lines) and  completely aligned ($S=1$, solid lines) packings. For $S=0$ the data along the $x$ and $y$ directions almost coincide. Here, $L(=16l, 32l, 64l, 128l)$ is the size of the system. $\nu = 4/3$ is the 2D correlation length percolation exponent~\cite{Stauffer1992}. \label{fig:f06}}
\end{figure}

Figure~\ref{fig:f06}a shows examples of the critical shell thickness $\delta_\text{c}$ versus the inverse systems size $1/L$ at different values of $\varphi$. Here, $L(=16l,32l,64l, 128l)$ is the size of the system. The data are presented for aspect ratio of $\varepsilon=4$ for completely disordered ($S=0$, dashed lines) and  completely aligned ($S=1$, solid lines) packings. Increase in $\varphi$ resulted in a decrease of $\delta_\text{c}$ and the minimum values of $\delta_\text{c}$ were observed at the jamming coverage ($\varphi=\varphi_\text{j}\approx 0.557$ for $\varepsilon=4$). For $S=0$, the data along the $x$ and $y$ directions almost coincide. However, for finite-sized aligned systems ($S\neq 0$), the value of $\delta_{c,y}$ always exceeded the value of $\delta_{c,x}$, and both these values exceeded the value $\delta_\text{c}$ for isotropic systems. Figure~\ref{fig:f06}b shows similar examples of the critical coverage $\varphi_\text{c}$ versus the value of $L^{-1/\nu}$ at different fixed values of shell thickness, $\delta$.  Here, $\nu=4/3$ is the 2D correlation length percolation exponent~\cite{Stauffer1992}. The data on the critical coverage $\varphi_\text{c}$ also demonstrated the presence of percolation anisotropy for the finite-sized aligned systems ($S\neq 0$). Similar percolation anisotropy was observed in finite-sized discrete systems with aligned rods ($k$-mers) and the finite size effects were also more pronounced for systems with aligned rods~\cite{Tarasevich2012,Tarasevich2016,Tarasevich2018}.
Thus, it can be concluded that anisotropies observed in the behavior of the critical shell thickness, $\delta_\text{c}$, and the critical coverage $\varphi_\text{c}$ are finite size scaling effects and that they disappear in the limit of $L\rightarrow\infty$. Moreover, the scaling behaviors of the value $\delta_\text{c}$ for completely disordered ($S=0$) and of the average value of $\delta_\text{c}=(\delta_{c,x}$+$\delta_{c,y})/2$ for aligned ($S\neq 0$) packings were fairly  insignificant for $L/l\geqslant 32$. Therefore, in the present work, the averaged values of $\delta_\text{c}$ and $\varphi_\text{c}$ in both directions were always used and all connectivity analysis tests were performed using $L/l = 32$.

\begin{figure}[!htbp]
	\centering
	\includegraphics[width=\columnwidth]{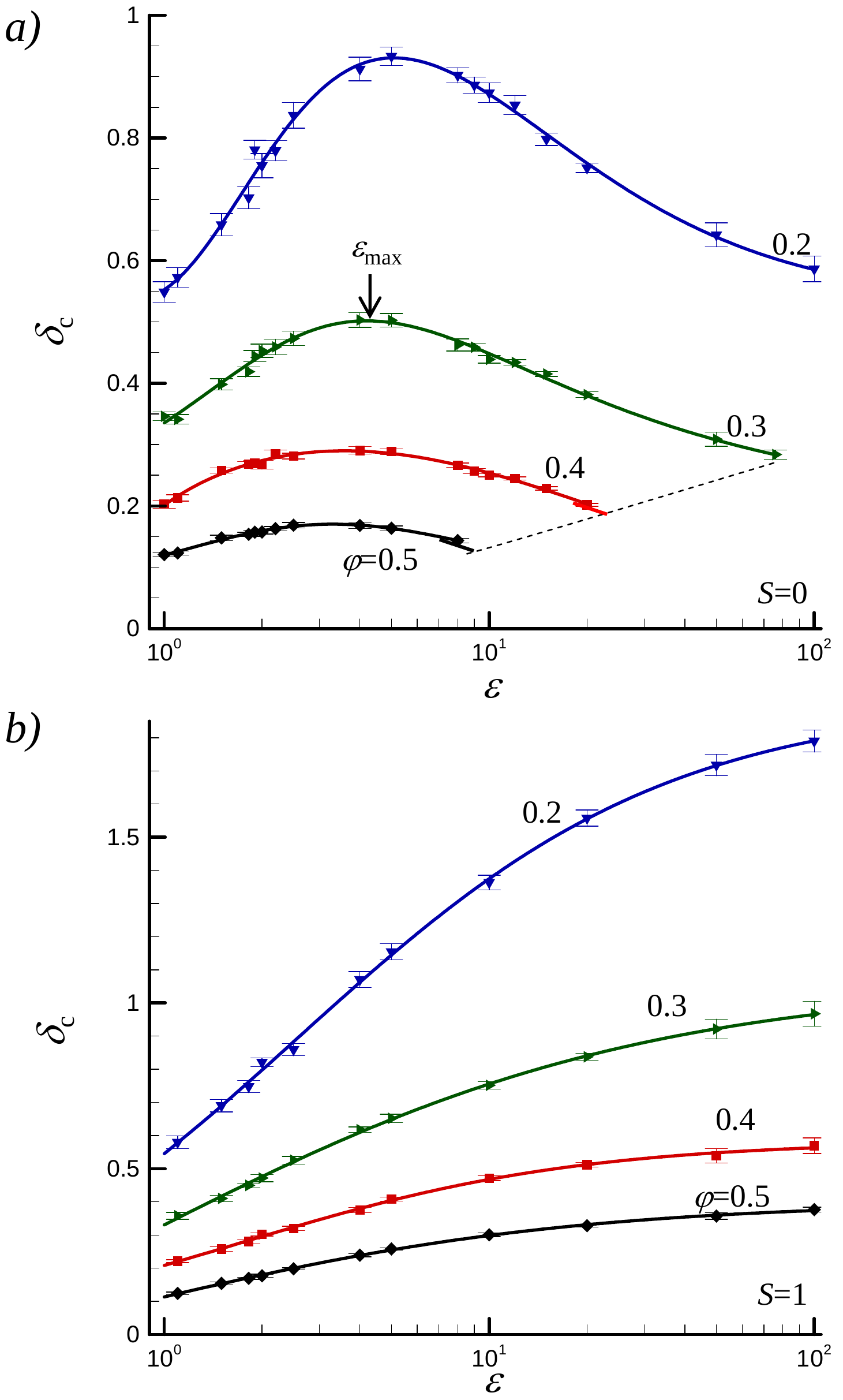}
\caption{Critical shell thickness $\delta_\text{c}$ versus the aspect ratio $\varepsilon$ at different coverages, $\varphi$ for completely disordered, $S=0$, (a) and  completely aligned, $S=1$, (b) packings. \label{fig:f07}}
\end{figure}

\begin{figure}[!htbp]
	\centering
	\includegraphics[width=\columnwidth]{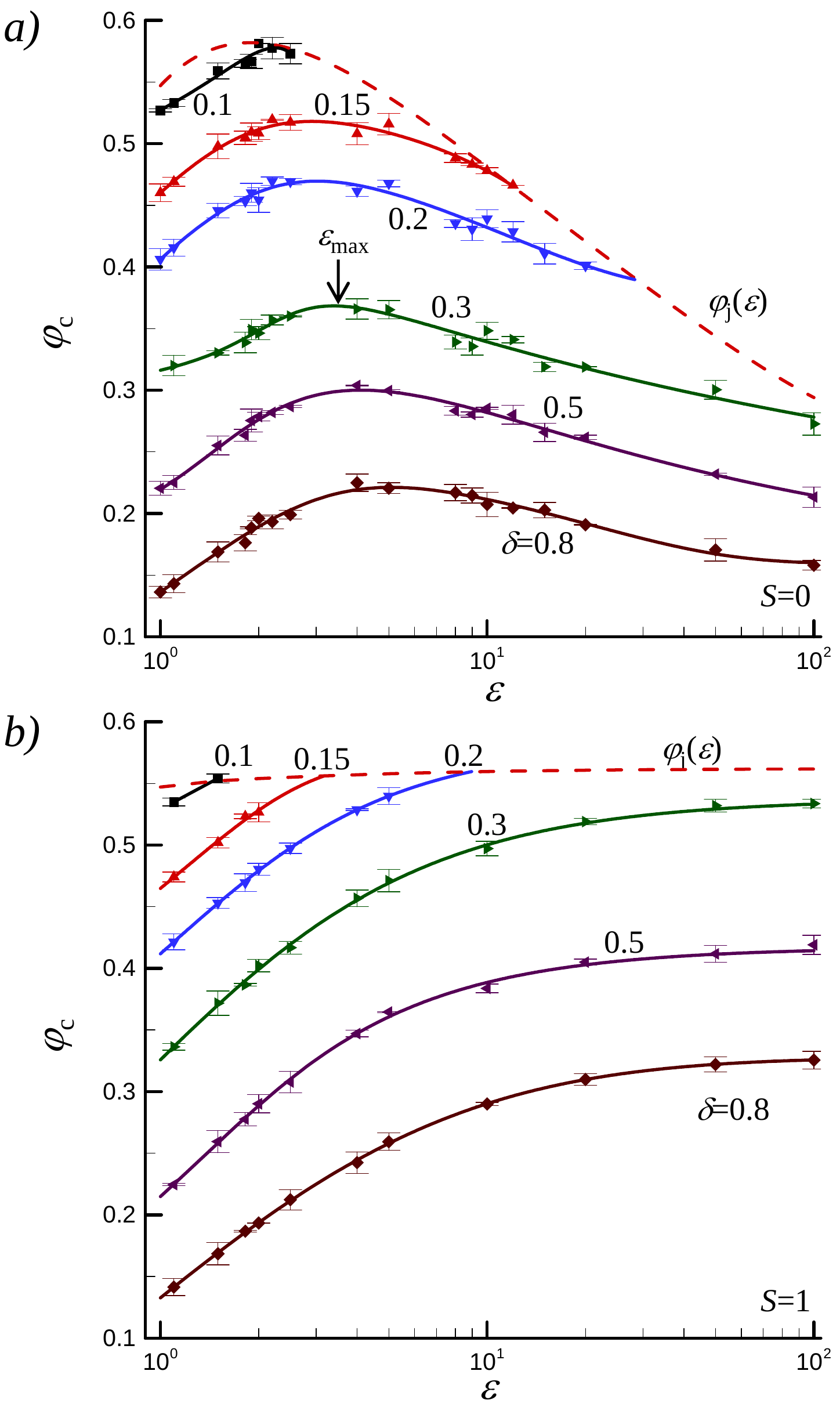}\\
	\caption{Critical coverage  $\varphi_\text{c}$ versus the aspect ratio $\varepsilon$ at different shell thickness, $\delta$, for completely disordered, $S=0$, (a) and  completely aligned, $S=1$, (b) packings.  \label{fig:f08}}
\end{figure}

Figure~\ref{fig:f07} and Fig.~\ref{fig:f08} demonstrate examples of the critical shell thickness $\delta_\text{c}$ (Fig.~\ref{fig:f07}), and the critical coverage $\varphi_\text{c}$ (Fig.~\ref{fig:f08}), versus the aspect ratio, $\varepsilon$, for completely disordered, $S=0$, (a); and  completely aligned, $S=1$, (b) packings. For completely disordered systems ($S=0$)  maximums on the $\delta_\text{c}(\varepsilon)$ (Fig.~\ref{fig:f07}a) and $\varphi_\text{c}(\varepsilon)$ (Fig.~\ref{fig:f08}a) curves, at some values of $\varepsilon_\text{max}$, were observed. The positions of these maximums were controlled by the values of $\varphi$ (Fig.~\ref{fig:f07}a) and $\delta$ (Fig.~\ref{fig:f08}a).

The observed maximums in the percolation characteristics $\delta_\text{c}$ and $\varphi_\text{c}$ may reflect the internal structure of the RSA packings of elongated particles. In particular,   maximums in the jamming coverage $\varphi_\text{j}$ versus the $\varepsilon$ dependencies were also observed for disordered packings and could be explained by the competition between the effects of orientational degrees of freedoms and excluded volume effects. The jamming limit decreased with $\varepsilon$~\cite{Lebovka2020b}, and for elongated particles in the vicinity of percolation packings, terminations of the curves $\delta_\text{c}(\varepsilon)$ (Fig.~\ref{fig:f07}a) and $\varphi_\text{c} (\varepsilon)$ (Fig.~\ref{fig:f08}a) at some critical values of $\varepsilon$ were observed.

These maximums became less pronounced for partially aligned systems, and they completely disappeared for completely aligned, $S=1$, packings (Fig.~\ref{fig:f07}b and Fig.~\ref{fig:f08}b). For the case of $S=1$, the values of $\delta_\text{c}$ (Fig.~\ref{fig:f07}b) and $\varphi_\text{c}$ (Fig.~\ref{fig:f08}b) grew with increasing values of $\varepsilon$, and for relatively small shell thickness, $\delta$, the termination of $\varphi_\text{c} (\varepsilon)$  was observed when the values of $\varphi_\text{c}$ exceed the jamming coverage, $\varphi_\text{j}$.

\subsection{Intrinsic conductivity}
The concept of intrinsic conductivity is useful for description of the behavior of the electrical conductivity in the limiting case of an infinitely diluted system. For randomly aligned and arbitrarily shaped particles with electrical conductivity $\sigma_p$ suspended in a continuous medium with electrical conductivity $\sigma_\text{m}$, the generalized Maxwell model gives the following virial expansion~\cite{Douglas1995,Garboczi1996}
\begin{equation}\label{eq:Garboczi1996}
\frac{\sigma}{\sigma_\text{m}} =1+[\sigma] \varphi+ \mathrm{O}(\varphi^2),
\end{equation}
where
\begin{equation}\label{eq:ic}
[\sigma] = \left.\frac{\mathrm{d}\ln\left(\sigma/\sigma_\text{m}\right)}{\mathrm{d}\varphi}\right|_{\varphi \to 0},
\end{equation}
is called the intrinsic conductivity, and $\varphi$ is the coverage concentration.

The value of the intrinsic conductivity $[\sigma]$ can depend upon the electrical conductivity contrast $\Delta=\sigma_p/\sigma_\text{m}$, the particle's aspect ratio, $\varepsilon$, the order parameter, $S$, and a spatial dimension.

\begin{figure}[!htbp]
	\centering	
\includegraphics[width=\columnwidth]{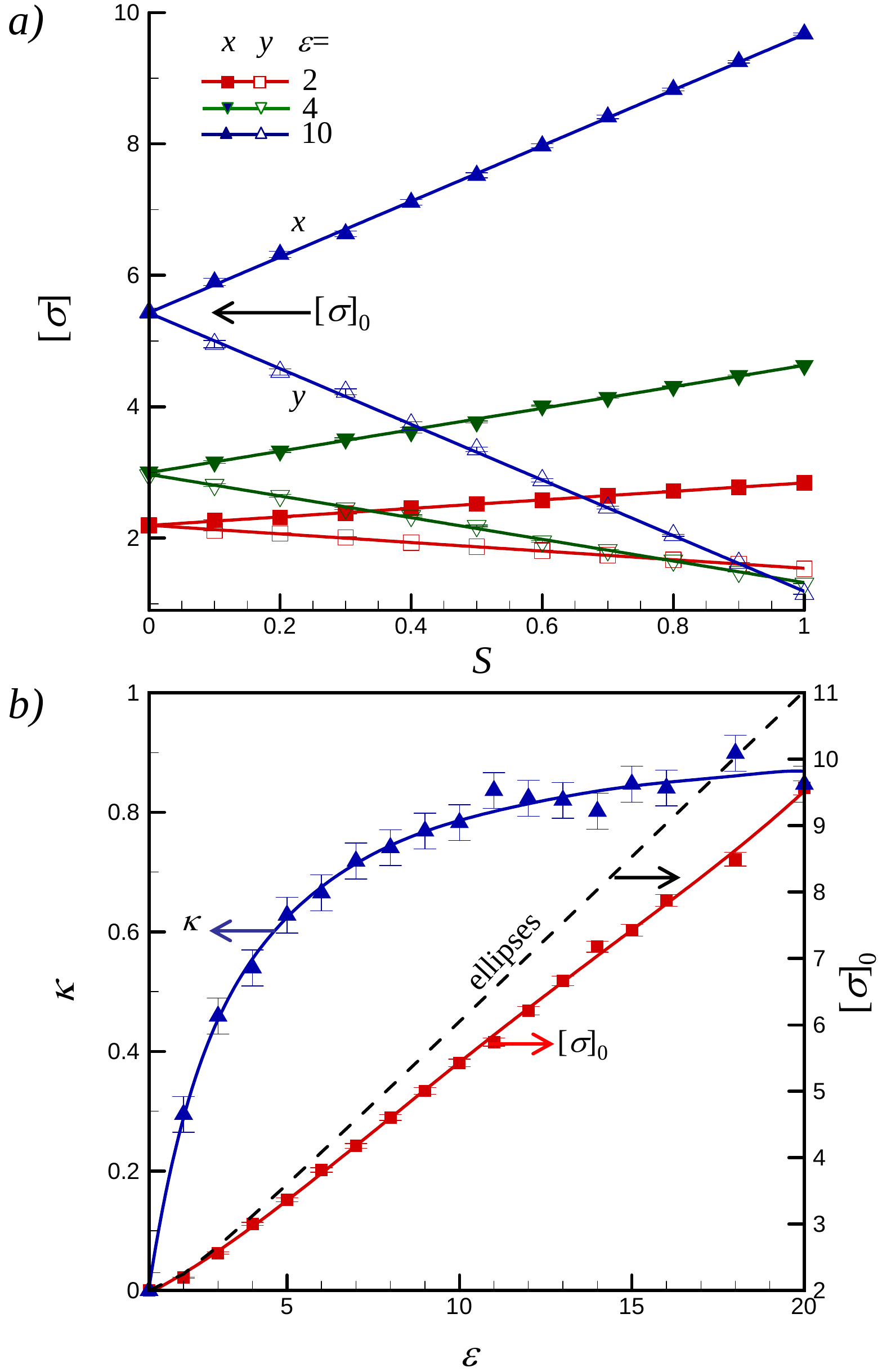}\\
	\caption{Examples of intrinsic conductivities $[\sigma]$ versus the order parameter, $S$. The data are presented along the $x$ and $y$ directions for discorectangles with aspect ratios $\varepsilon=2,4,10$ (a). Dependencies of the parameters $[\sigma]_0$, $\kappa$ (See Eq.~\ref{eq:icS}) versus $\varepsilon$ (b).\label{fig:f04}}.
\end{figure}

Figure~\ref{fig:f04}a demonstrates examples of intrinsic conductivities $[\sigma]$ versus the order parameter, $S$. The data are presented in the $x$ and $y$ directions for discorectangles with different aspect ratios $\varepsilon$. These dependencies were obtained using a mesh parameter of $m=4096$ over 1000 independent runs. The observed $[\sigma]$ versus $S$ relationships were almost  linear:
\begin{equation}\label{eq:icS}
[\sigma]=[\sigma]_0 (1 \pm\kappa S),
\end{equation}
where $[\sigma]_0$ is the intrinsic conductivity for the isotropic system with $S=0$,  $\kappa$ is the anisotropy coefficient, and the signs $+$ or $-$ correspond to the $x$ or $y$ directions, respectively.

Therefore, the intrinsic conductivity $[\sigma]_x$ along alignment direction $x$ exceeded  value  $[\sigma]_y$ in the perpendicular direction hence symmetric behavior with the same anisotropy coefficients $\kappa$ was observed.

Figure~\ref{fig:f04}b presents values of $[\sigma]_0$ and $\kappa$ versus the aspect ratio $\varepsilon$. The intrinsic conductivity for the isotropic system  $[\sigma]_0$ increased with $\varepsilon$. Note, that similar behavior has been predicted theoretically for randomly aligned ellipses ($S=0$)~\cite{Douglas1995,Garboczi1996}
\begin{equation}\label{eq:Sangani1990sigma}
[\sigma]=\frac{(\Delta^2-1)(1+\varepsilon)^2}{2(1+\varepsilon\Delta)(\Delta+\varepsilon)}.
\end{equation}
For  $\Delta \gg 1$, this equation gives (see dashed line in Fig.~\ref{fig:f04}b)
\begin{equation}\label{eq:Elliptical1}
[\sigma] = 1 + \frac{1}{2} \left( \varepsilon + \frac{1}{\varepsilon} \right).
\end{equation}
The anisotropy coefficient $\kappa$ also increased with $\varepsilon$. It presumably  tend to the unit in the limit $\varepsilon\gg 1$.

\begin{figure}[!htbp]
	\centering	
\includegraphics[width=\columnwidth]{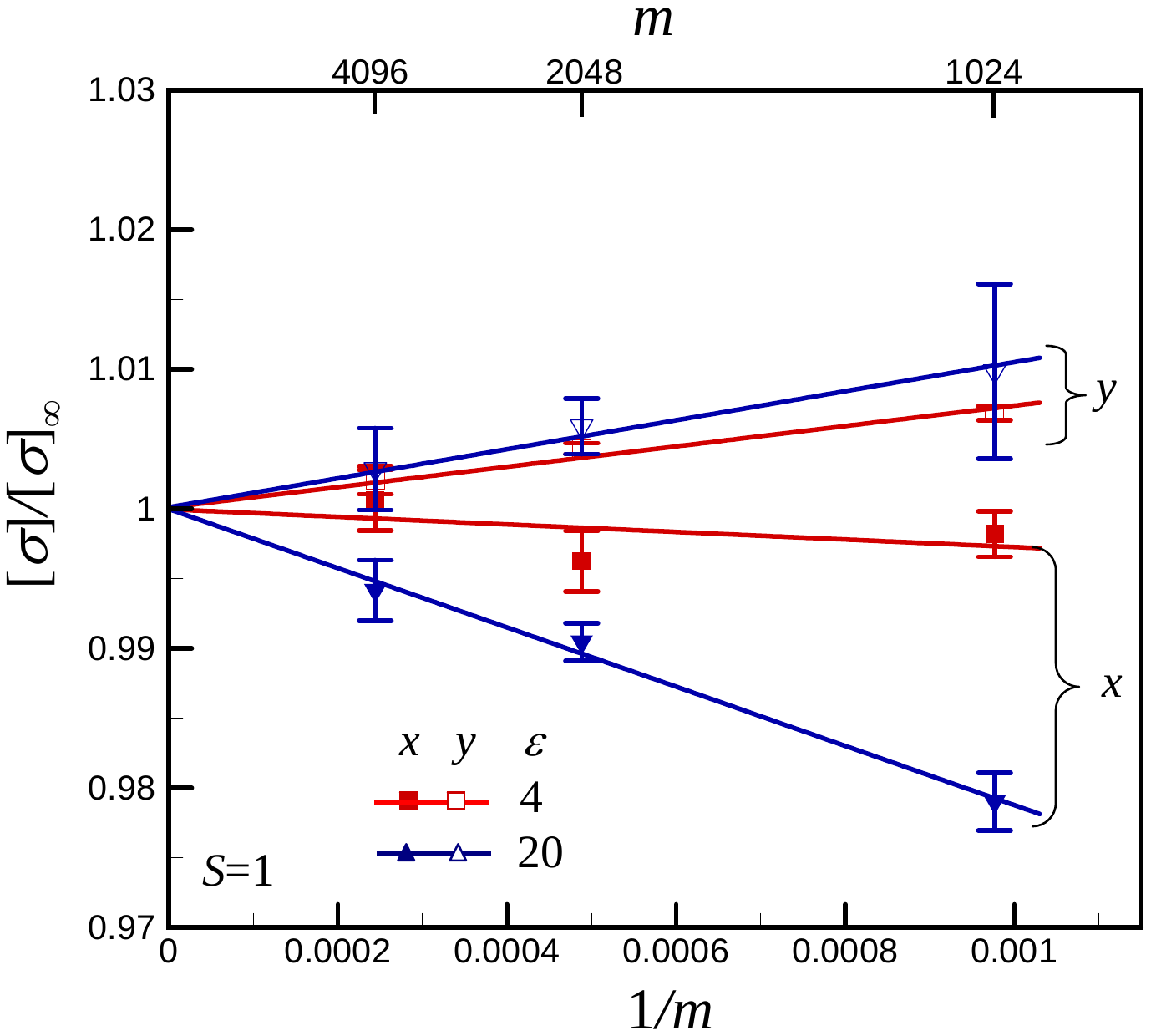}\\
	\caption{Normalized intrinsic conductivity $[\sigma]/[\sigma]_\infty$ in the $x$ and $y$ directions versus the inverse mesh size $1/m$ at different aspect ratios $\varepsilon=4,20$, and $S=1$. Here, the value of $[\sigma]_\infty $ corresponds to the intrinsic conductivity in the limit of $m\to\infty$. \label{fig:f05}}
\end{figure}

Figure~\ref{fig:f05} illustrates the effect of mesh size $m$ on the precision of $[\sigma]$ determination at two values of the aspect ratio $\varepsilon$. The observed $[\sigma]$ versus the inverse mesh size $1/m$ were almost linear: $[\sigma]=[\sigma]_\infty (1 +a/m),$
where $[\sigma]_\infty $ and $a$ are the fitting parameters. The data evidenced that estimation errors of $[\sigma]$  increased with increasing value of $\varepsilon$ reaching about $2\%$ for $\varepsilon=20$ and $m=1024$.

\subsection{Electrical conductivity\label{subsec:conductivity}}

\begin{figure}[!htbp]
	\centering
	\includegraphics[width=\columnwidth]{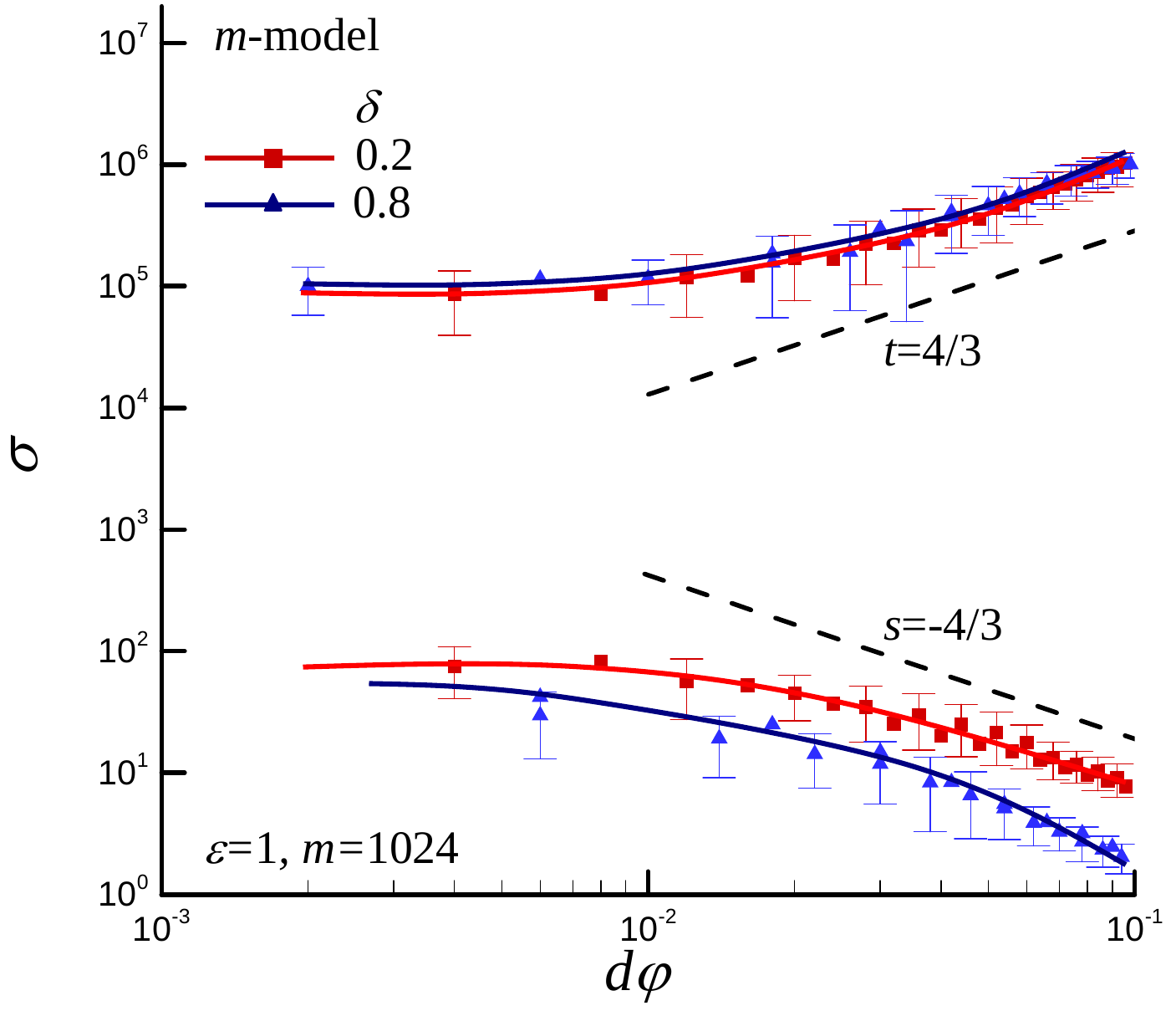}
\caption{Electrical conductivity,  $\sigma$, versus the difference, $\mathrm{d}\varphi=\left|\varphi-\varphi_\sigma\right|$, for RSA packings of disks ($\varepsilon=1$) at different shell thicknesses $\delta$. Here, the value of $\varphi_\sigma$ was identified from the concentration at the percolation jump for each independent run, with calculations being carried out using a mesh size of $m=1024$. Dashed lines corresponds to the classical exponents $s=t\approx 4/3$~\cite{Stauffer1992}.\label{fig:f09}}
\end{figure}

For each independent run the electrical conductivity $\sigma$ displayed a jump at some percolation concentration $\varphi_\sigma$. Figure~\ref{fig:f09} presents $\sigma$, versus the difference, $\mathrm{d}\varphi=\left|\varphi-\varphi_\sigma\right|$, for RSA packings of disks ($\varepsilon=1$) at the different shell thicknesses, $\delta=0.2$ and $\delta=0.8$.

\begin{figure}[!htbp]
	\centering
	\includegraphics[width=\columnwidth]{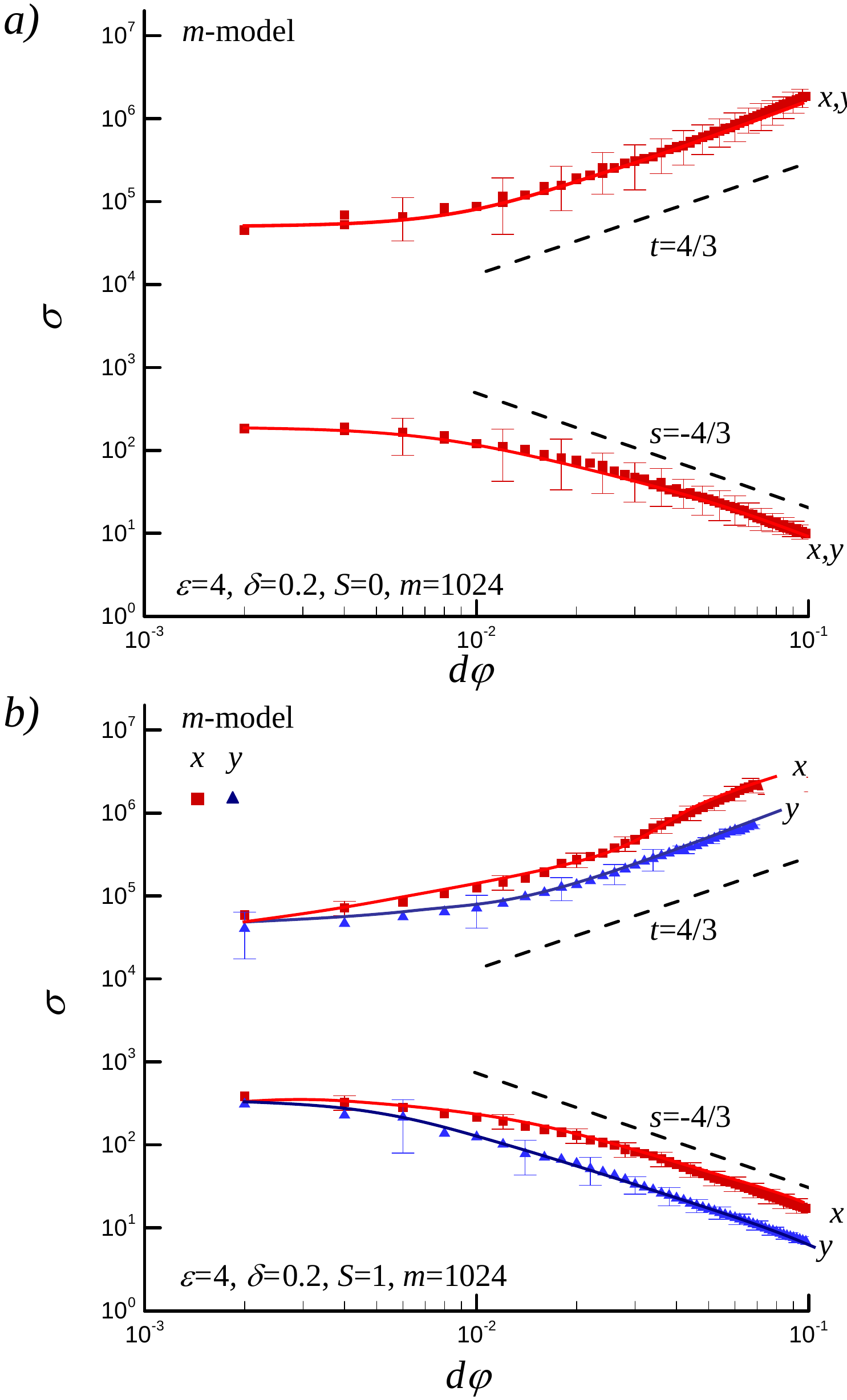}\\
	\caption{Electrical conductivity,  $\sigma$, versus the difference, $\left|\varphi-\varphi_\sigma\right|$, for RSA packings of discorectangles with different values of aspect ratios, $\varepsilon$, for a fixed shell thickness of $\delta=0.2$ for completely disordered, $S=0$, (a) and  completely aligned, $S=1$, (b) packings. Here, the value of $\varphi_\sigma$ was identified from the concentration at the percolation jump for each independent run, with the calculations being performed using a mesh size of $m=1024$. Dashed lines corresponds to the classical exponents $s=t\approx 4/3$~\cite{Stauffer1992}.\label{fig:f10}}
\end{figure}

In order to check for the possible non-universality of the percolation exponents, the critical conductivity indexes $s$ and $t$ were estimated from the scaling relations for the electrical conductivities just below, $\sigma \propto (\mathrm{d}\varphi)^{-s}$, and above, $\sigma \propto (\mathrm{d}\varphi)^{t}$, the percolation threshold~\cite{Stauffer1992}. The classical values for 2D percolation are $s=t\approx 4/3$. Obtained data evidenced the satisfactory correspondence of the percolation exponents to the classical universality. Below the percolation threshold the difference between the curves for $\delta=0.2$ and $\delta=0.8$ evidently reflected the effects of the shell thickness on the value of $\varphi_\sigma$. Above the percolation threshold, such effects were insignificant. Figure~\ref{fig:f10} compares  $\sigma$, versus the difference, $\mathrm{d}\varphi=\left|\varphi-\varphi_\sigma\right|$, dependencies, for RSA packings of discorectangles ($\varepsilon=4$) at a fixed value of $\delta=0.2$ for completely disordered, $S=0$, (a); and  completely aligned, $S=1$, (b) packings. For aligned packings, a significant anisotropy  in the electrical conductivity was observed and the values along the alignment direction, $\sigma_x$, significantly exceeded the values in the perpendicular direction, $\sigma_y$. Importantly, the obtained data for the mesh sizes of $m=1024$ and $m=2048$ were approximately the same within data errors.

\begin{figure}[!htbp]
	\centering
	\includegraphics[width=\columnwidth]{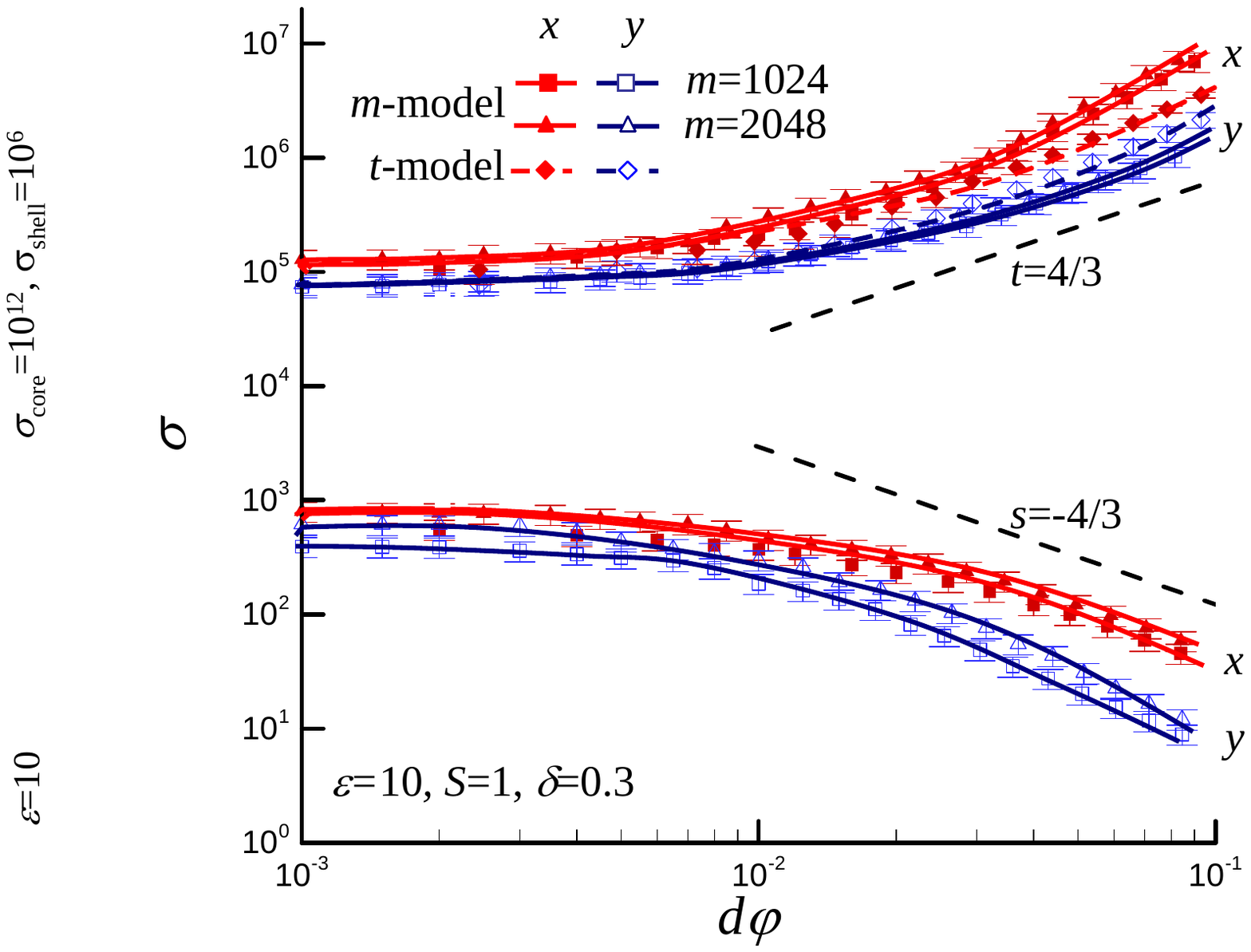}\\
	\caption{Electrical conductivity,  $\sigma$, versus the difference, $\left|\varphi-\varphi_\sigma\right|$. The data are presented for discorectangles with an aspect ratio $\varepsilon=10$ and a shell thickness of $\delta=0.3$, for completely aligned, $S=1$, RSA packings. Here, the value of $\varphi_\sigma$ was identified from the concentration at the  percolation jump for each independent run, with the calculations being performed using mesh sizes of $m=1024$ and $m=2048$. Moreover, above the percolation threshold, the electrical conductivity obtained within t-model is also presented. Dashed lines corresponds to the classical exponents $s=t\approx 4/3$~\cite{Stauffer1992}.\label{fig:f11}}
\end{figure}

Figure~\ref{fig:f11} compares the electrical conductivity, $\sigma$, versus the difference, $\mathrm{d}\varphi=\left|\varphi-\varphi_\sigma\right|$ for fairly long discorectangles ($\varepsilon=10$). The data are presented at a fixed value of $\delta=0.3$ for completely aligned ($S=1$) RSA packings at two values of $m$. The observed behavior for $\varepsilon=10$ was similar to that seen with $\varepsilon=4$ (Fig.~\ref{fig:f10}b). Above the percolation threshold ($\varphi>\varphi_\sigma$) the effect of $m$ was insignificant. However, below percolation threshold ($\varphi<\varphi_\sigma$) the electrical conductivities estimated at $m=1024$ were systematically smaller compared to those estimated at $m=2048$. Above the percolation threshold, the electrical conductivities obtained within m-model and t-model demonstrate similar behavior.

\begin{figure}[!htbp]
	\centering
	\includegraphics[width=\columnwidth]{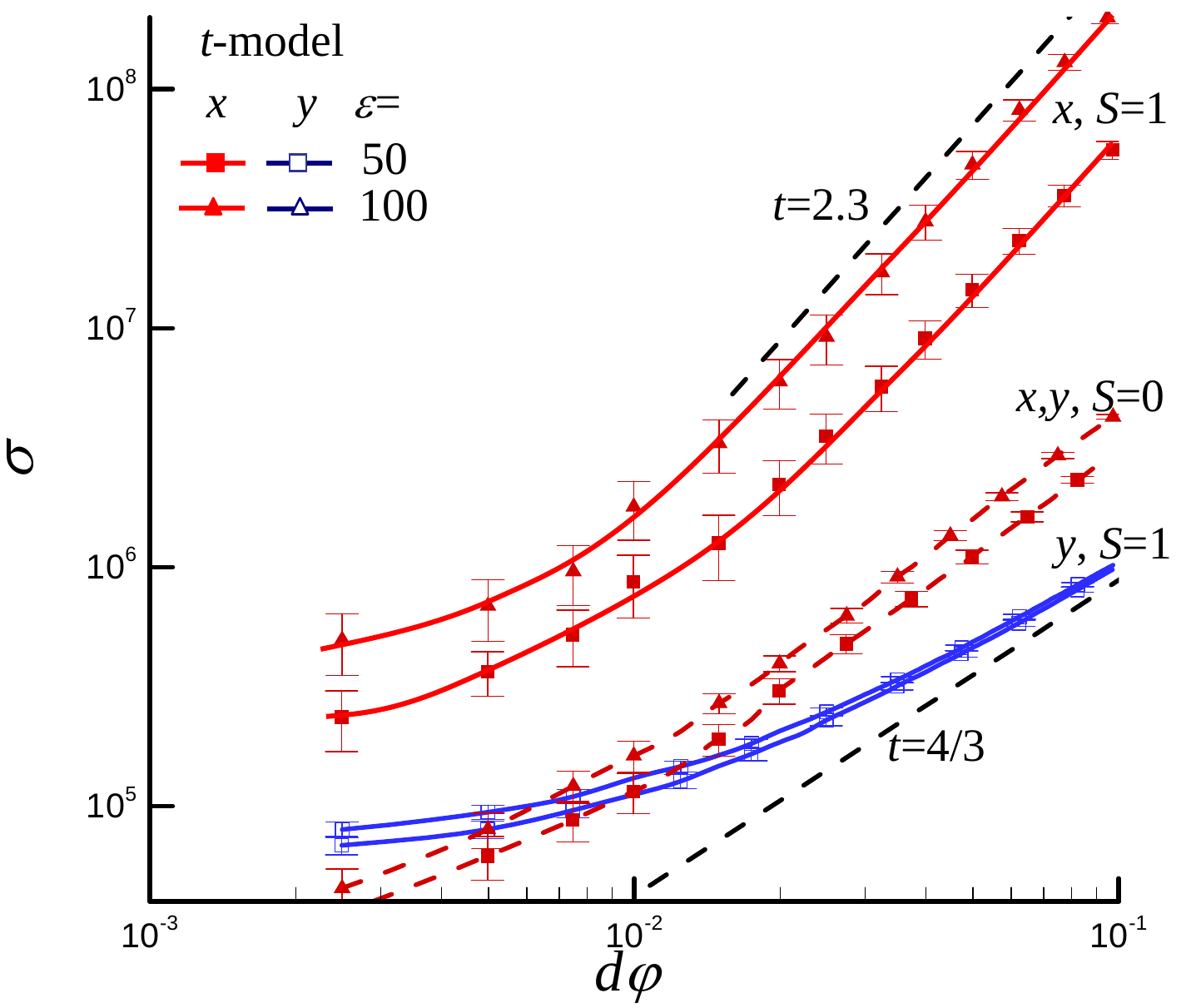}\\
	\caption{Electrical conductivity,  $\sigma$, versus the difference, $\left|\varphi-\varphi_\sigma\right|$. The data are presented for discorectangles with the aspect ratios $\varepsilon=50,100$ and a shell thickness of $\delta=0.8$, for completely disordered, $S=0$, and completely aligned, $S=1$, RSA packings. Here, the value of $\varphi_\sigma$ was identified from the concentration at the  percolation jump for each independent run. Dashed lines corresponds to the exponents $t\approx 2.3$ and $t\approx 4/3$~\cite{Stauffer1992}.\label{fig:f12}}
\end{figure}

Finally, Fig.~\ref{fig:f12} compares the electrical conductivity, $\sigma$, versus the difference, $\mathrm{d}\varphi=\left|\varphi-\varphi_\sigma\right|$ for long discorectangles ($\varepsilon=50,100$). The data are presented at a fixed value of $\delta=0.8$ for completely disordered, $S=0$, and completely aligned ($S=1$) RSA packings. The results have been obtain within t-model.

For completely disordered systems ($S=0$), obtained data evidenced the correspondence of the percolation exponents to the classical universality, $t\approx 4/3$. However, for completely aligned RSA packings ($S=1$), significant anisotropy in electrical conductivity and deviations from the classical universality were observed. In the direction of alignment, $x$,  the exponents $t\approx 2.3$ were observed for the both $\varepsilon =50$ and $\varepsilon =100$ values. In the perpendicular direction, $y$, the exponents closer to the classical universality value $t\approx 4/3$ were observed. The similar non-universal values of the critical conductivity exponents were observed for systems of penetrable sticks and nanowires~\cite{Keblinski2004,Li2010,Zevzelj2012,Han2018,Kim2019,Ponzoni2019,Fata2020}. Particularly, $t$ transitions from $\approx 1$ to $\approx 2$ was observed in nanowire-to-junction resistance dominated networks~\cite{Fata2020}. The effects of widthless stick alignment on the percolation critical exponents were also observed. In our case, for impenetrable very elongated particles with core--shell structure, the change in the critical exponent may reflect the changes in the morphology of conducting paths in the networks with a change in coverage.

\section{Conclusion\label{sec:conclusion}}

Numerical studies of two-dimensional RSA deposition of aligned discorectangles on a plane were carried out. The resulting partial ordering was characterized by the order parameter $S$, with $S=0$ for random orientation of the particles and $S=1$ for completely aligned particles in the horizontal direction $x$. Analysis of connectivity was performed assuming a core--shell structure of the particles. The values of the aspect ratio, $\varepsilon$, and order parameter, $S$, significantly affected the structures of the packings, the formation of long-range connectivity and of the behavior of the electrical conductivity. The observed effects probably reflect the competition between the particles' orientational degrees of freedom and the excluded volume effects~\cite{Chaikin2006}. For aligned systems, different anisotropies in intrinsic conductivity, long range connectivity, and the behavior of electrical conductivity were observed. For example, a significant anisotropy in electrical conductivity was observed and the values in the alignment direction, $\sigma_x$, were larger than the values in the perpendicular direction, $\sigma_y$.  For aligned finite-size systems, the percolation thresholds in the $x$ and $y$ directions were different.  However, these differences disappeared in the limit of infinitely large systems.

\begin{acknowledgments}
We are thankful to I.V.Vodolazskaya for our stimulating discussions and to A.G.Gorkun for technical assistance.
We acknowledge funding from the National research foundation of Ukraine, Grant No.~2020.02/0138 (M.O.T., N.V.V.), the National Academy of Sciences of Ukraine, Project Nos.~7/9/3-f-4-1230-2020,~0120U100226 and~0120U102372/20-N (N.I.L.), and funding from the Foundation for the Advancement of Theoretical Physics and Mathematics ``BASIS'', Grant No. 20-1-1-8-1. (Y.Y.T. and A.V.E.).
\end{acknowledgments}

\appendix
\section{Description of m-model for calculation of electrical conductivity\label{sec:appendixmmodel}}

\begin{figure}[!htbp]
	\centering
	\includegraphics[width=\columnwidth]{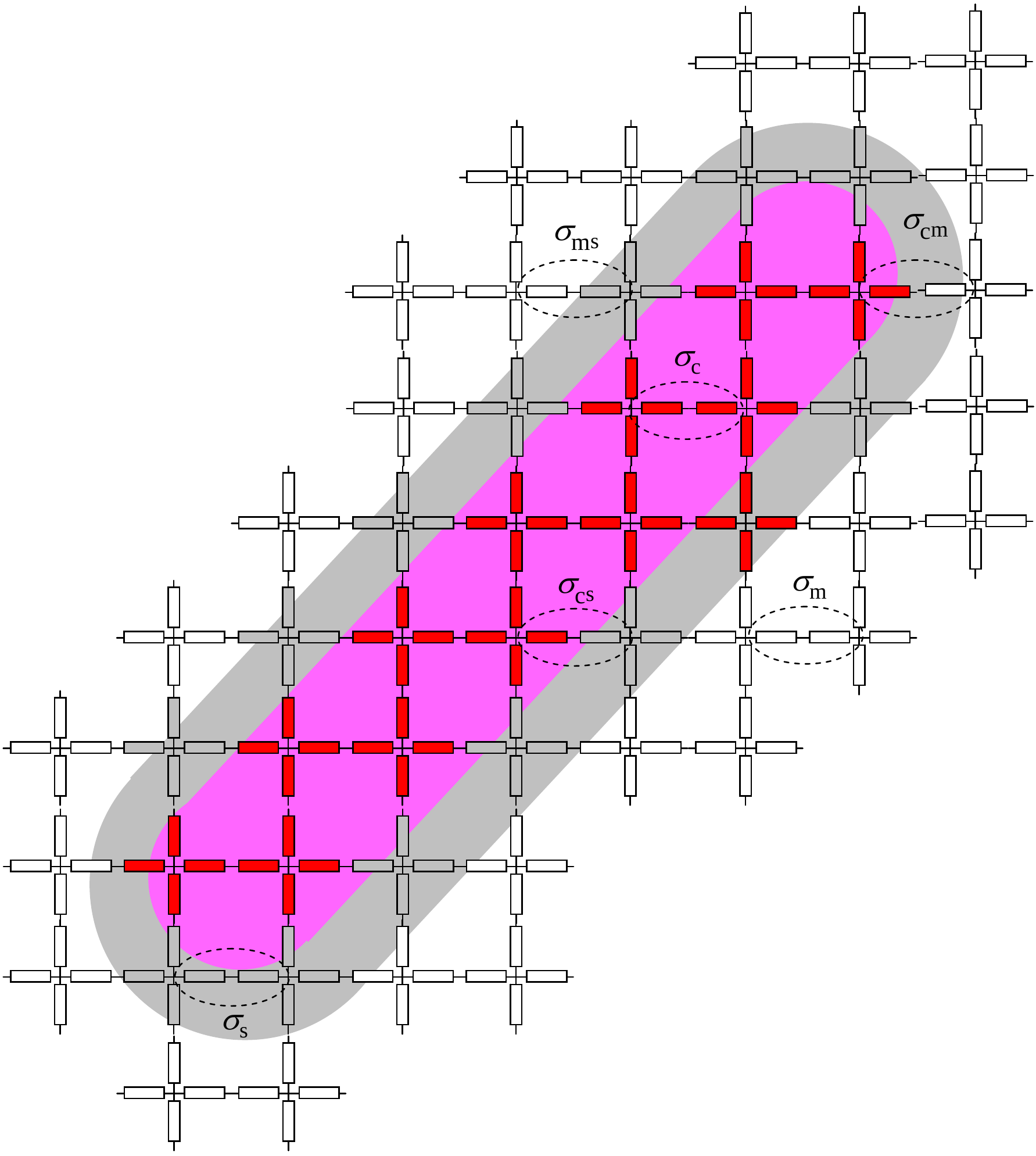}
	\caption{ Representation of the mesh square lattice with deposited discorectangle. The centers of the mesh cells were located at the cores, shells, or pores parts. Each cell was associated with a set of four resistors. The electrical conductivities of the whole bonds between two similar sites were $\sigma_\text{c}$, $\sigma_\text{s}$ and $\sigma_\text{m}$. For bonds located between different sites, there are only three possible combinations of the electrical conductivities of the entire bonds between core and shell sites, $\sigma_\text{cs}$, pore and shell sites, $\sigma_\text{ms}$, and core and pore sites, $\sigma_\text{cm}$.
 \label{fig:f03A}}
\end{figure}

The mesh cells (sites) with centers located at the core, shell, or pore parts were assumed to have electrical conductivities of $\sigma_\text{c}$, $\sigma_\text{s}$, and $\sigma_\text{m}$, respectively. Each cell was associated with a set of four resistors. The electrical conductivities of the whole bonds between two similar sites were calculated as $\sigma_\text{c}$, $\sigma_\text{s}$, and $\sigma_\text{m}$ when the both sites were located at the core, shell, or pore parts, respectively (Fig.~\ref{fig:f03A}). For bonds located between different sites, there are only three possible combinations of the electrical conductivities of the entire bonds between core and shell sites, $\sigma_\text{cs}$, pore and shell sites, $\sigma_\text{ms}$, and core and pore sites, $\sigma_\text{cm}$. The electrical conductivities of the entire bonds were calculated as
$\sigma_\text{cs}=2\sigma_\text{c}\sigma_\text{s}/(\sigma_\text{c}+\sigma_\text{s})$ (between core and shell sites),
$\sigma_\text{ms}=2\sigma_\text{m}\sigma_\text{s}/(\sigma_\text{m}+\sigma_\text{s})$ (between pore and shell sites), and
$\sigma_\text{cm}=2\sigma_\text{c}\sigma_\text{m}/(\sigma_\text{c}+\sigma_\text{m})$ (between core and pore sites).

\section{The way of calculating the area of intersection of the two discorectangles (stadia)\label{sec:appendix}}

\begin{figure}
  \centering
  \includegraphics[width=\columnwidth]{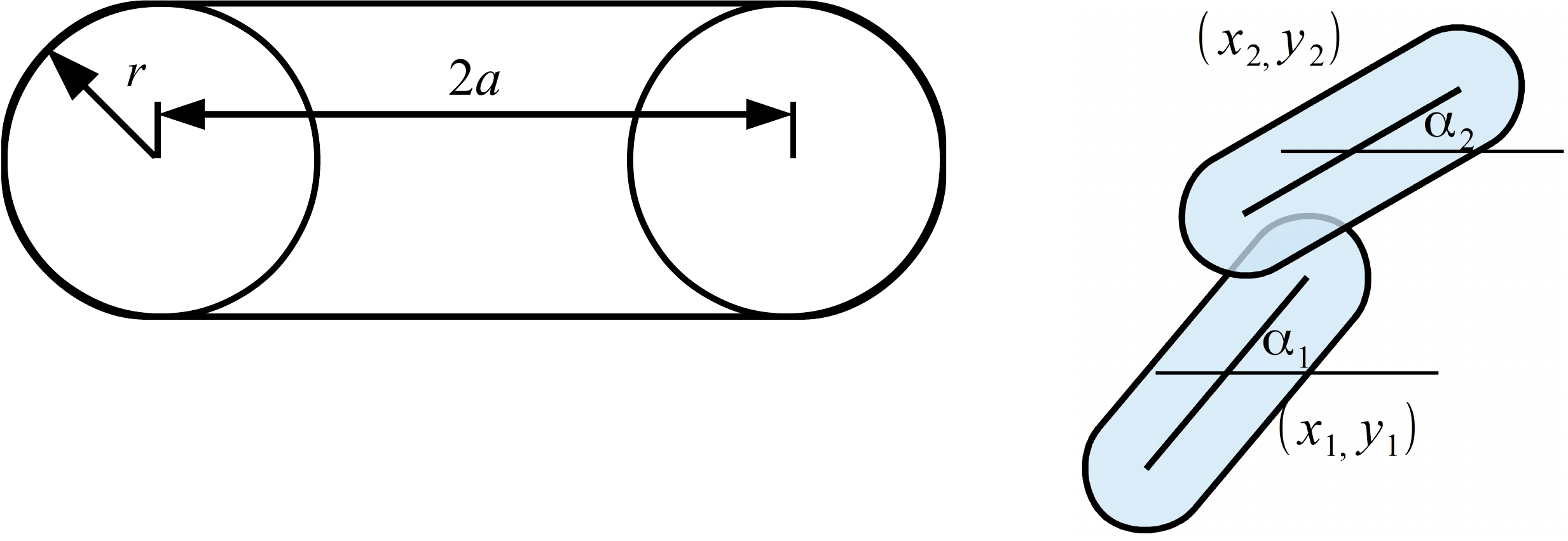}
  \caption{Explanation of the notation used. $(x_1, y_1) \in\mathbb{R}^2$, $\alpha_1\in[-\pi/2,\pi/2)$, $(x_2, y_2)\in\mathbb{R}^2$, $\alpha_2\in[-\pi/2,\pi/2)$. ($d=2r$, $l = 2(r+a)$.)\label{fig:AppendixFig}}
\end{figure}

Calculating the area of intersection of the two discorectangles (stadia) is used the notation explained in Fig~\ref{fig:AppendixFig}.

Functions of the boundaries are presented in Table~\ref{tab:boundaries}.
\begin{table*}
\caption{Functions of the boundaries.\label{tab:boundaries}}
\begin{ruledtabular}
\begin{tabular}{l|l|l}
  $x$ & $y(x)$ & Equations in canonical form\\
  \hline
  \multicolumn{3}{c}{Function of the upper boundary of the $i$-th stadium $F_+^i$, when $\alpha_i\in(-\pi/2,\pi/2)$:}\\
  $x_0\left(F_+^i\right)  = x_i-a \cos\alpha_i-r$ &  &\\
   & $\sqrt{r^2-(x-x_i+a \cos\alpha_i)^2} + y_i - a \sin\alpha_i$  & $(x- x_i + a \cos\alpha_i)^2+(y - y_i + a \sin\alpha_i)^2=r^2$ \\
  $x_1\left(F_+^i\right)  = x_i-a \cos\alpha_i-r \sin\alpha_i$ & & \\
   & $(x+r \sin\alpha_i-x_i)\tan\alpha_i + y_i +r \cos\alpha_i$ & $  x\sin\alpha_i -  y \cos\alpha_i + r -  x_i \sin\alpha_i +   y_i \cos\alpha_i =0$ \\
  $x_2\left(F_+^i\right)  = x_i+a \cos\alpha_i-r \sin\alpha_i$ &  &\\
   & $\sqrt{r^2-(x-x_i-a \cos\alpha_i)^2} + y_i + a \sin\alpha_i$ & $(x - x_i - a \cos\alpha_i)^2+(y- y_i - a \sin\alpha_i)^2=r^2$ \\
   $x_3\left(F_+^i\right)  = x_i+a \cos\alpha_i+r$ &  &\\
  \hline
  \multicolumn{3}{c}{Function of the upper boundary of the $i$-th stadium $F_+^i$, when $\alpha_i=-\pi/2$:}\\
  $x_0\left(F_+^i\right)  = x_i-r$ & &\\
&$\sqrt{r^2-(x-x_i)^2} + y_i + a$ & $(x-x_i)^2 + (y - y_i - a)^2= r^2$ \\
$x_1\left(F_+^i\right)  = x_i+r$ & &\\
  \hline
  \multicolumn{3}{c}{Function of the lower boundary of the $i$-th stadium $F_-^i$, when $\alpha_i\in(-\pi/2,\pi/2)$:}\\
 $x_0\left(F_-^i\right)  = x_i-a \cos\alpha_i-r$ &  &\\
& $-\sqrt{r^2-(x-x_i+a \cos\alpha_i)^2} + y_i - a \sin\alpha_i$  & $(x-x_i + a \cos\alpha_i)^2+(y - y_i + a \sin\alpha_i)^2=r^2$ \\
$x_1\left(F_-^i\right)  = x_i-a \cos\alpha_i+r \sin\alpha_i$ & &\\
& $(x-r \sin\alpha_i-x_i)\tan\alpha_i + y_i -r \cos\alpha_i$ & $  x\sin\alpha_i -  y \cos\alpha_i - r -  x_i \sin\alpha_i +   y_i \cos\alpha_i =0$ \\
$x_2\left(F_-^i\right)  = x_i+a \cos\alpha_i+r \sin\alpha_i$ & &\\
& $-\sqrt{r^2-(x-x_i-a \cos\alpha_i)^2} + y_i + a \sin\alpha_i$  & $(x - x_i - a \cos\alpha_i)^2+(y - y_i - a \sin\alpha_i)^2=r^2$ \\
$x_3\left(F_-^i\right)  = x_i+a \cos\alpha_i+r$ & &\\
  \hline
  \multicolumn{3}{c}{Function of the lower boundary of the $i$-th stadium $F_-^i$, when $\alpha_i=-\pi/2$ :}\\
$x_0\left(F_-^i\right)  = x_i-r$ & &\\
& $-\sqrt{r^2-(x-x_i)^2} + y_i - a$ & $(x-x_i)^2+(y - y_i + a)^2= r^2$ \\
$x_1\left(F_-^i\right)  = x_i+r$ & &\\
  \end{tabular}
\end{ruledtabular}
\end{table*}

Intersection of a circle $(x-x_1)^2 + (y-y_1)^2 = r^2$
and a line $Ax + By + C = 0$ is
$$
d = \sqrt{r^2 - \frac{(Ax_1+By_1+C)^2}{A^2+B^2}}.
$$
If the radical expression $\le 0$, then there are no intersections or there is only tangency, and we return an empty set of additional points. Otherwise, the intersection points are
$$
x = \frac{B^2 x_1-ABy_1-CA}{A^2+B^2} \pm \frac{dB}{\sqrt{A^2+B^2}}.
$$

Intersection of the two circles
$$
(x-x_1)^2 + (y-y_1)^2 = r^2
$$
and
$$
(x-x_2)^2 + (y-y_2)^2 = r^2
$$
$$
D = (x_2-x_1)^2 + (y_2-y_1)^2
$$
is the square of the distance between the centers of the circles.

If $D\ge 2r$, then there are no intersections or the circles are tangent, and we return an empty set of additional points.

Otherwise, the intersection points are
$$
x = \frac{x_1+x_2}{2} \pm (y_2 - y_1)\sqrt{\frac{r^2}{D} - \frac{1}{4}}.
$$

Intersection of the two lines $A_1	x + B_1 y + C_1 = 0$ and $A_2 x + B_2 y + C_2 = 0$
is $D = A_1B_2-A_2B_1.$

If $D=0$, then there are no intersections or the lines coincide, and we return an empty set of additional points.

Otherwise, the intersection points are
$$
x = \frac{B_1C_2-B_2C_1}{D}.
$$

$$
S = \int \left[\min\left(F_+^1, F_+^2\right) - \max\left(F_-^1, F_-^2\right) \right]_+
$$
is the master equation, where $(x)_+ = \max(x, 0)$.

We define the function $\min\left(F_+^1, F_+^2\right).$

We need to take the two lists (already ordered ascending)
$x_0\left(F_+^1\right), \dots , x_k\left(F_+^1\right), \; (k=1,3)$ and $x_0\left(F_+^2\right), \dots , x_m\left(F_+^2\right), \; (m=1,3)$
combine them into one (ascending list) and remove from this list all values smaller than $\max\left[ x_0\left(F_+^1\right), x_0\left(F_+^2\right) \right]$ and all values larger than $\min\left[ x_k\left(F_+^1\right), x_m\left(F_+^2\right) \right]$.

Let's get an ordered list $t_0, \dots , t_n$. For each $[t_i,t_{i+1}]$ ($0\le i<n$), the explicit analytical form of functions $F_+^1, F_+^2$ is uniquely determined.
To determine functions on an interval, it is enough to look in which interval of the domain of definition of functions $F_+^1, F_+^2$ lies the middle of this segment. Then we determine the intersection points, if any. If these intersection points are in this interval, then we add them, but we do not change the analytical functions.

After the procedure of dividing the region by intersection points, we can set the function $\min\left(F_+^1, F_+^2\right)$. On each interval of two functions, we leave only one, the value of which is less in the middle of the interval. We define the functions $\max\left(F_-^1, F_-^2\right)$ and $\left[\min\left(F_+^1, F_+^2\right) - \max\left(F_-^1, F_-^2\right) \right]_+$.

$$
S = \int \left[\min\left(F_+^1, F_+^2\right) - \max\left(F_-^1, F_-^2\right) \right]_+
$$
is the result of taking a definite integral over each interval and summing the results.
\begin{multline*}
\int{\left( \sqrt{{{r}^{2}}-{{(x-a)}^{2}}}+b \right)}\, \mathrm{d}x=\frac{x-a}{2}\sqrt{{{r}^{2}}-{{(x-a)}^{2}}}\\
+\frac{{{r}^{2}}}{2}\arctan \left( \frac{x-a}{\sqrt{{{r}^{2}}-{{(x-a)}^{2}}}} \right)+bx,
\end{multline*}
\begin{multline*}
  \int{\left( -\sqrt{{{r}^{2}}-{{(x-a)}^{2}}}+b \right)}\, \mathrm{d}x =-\frac{x-a}{2}\sqrt{{{r}^{2}}-{{(x-a)}^{2}}}\\
-\frac{{{r}^{2}}}{2}\arctan \left( \frac{x-a}{\sqrt{{{r}^{2}}-{{(x-a)}^{2}}}} \right)+bx,
\end{multline*}
$$
\int (ax + b) \, \mathrm{d}x = \frac{ax^2}{2} + bx.
$$

\bibliography{Lebovka2020Percolation}  

\end{document}